\documentclass[epsfig,a4paper,12pt]{article}
\usepackage{epsfig}
\usepackage{amstex}

\begin{document}
{\thispagestyle{empty}
\setcounter{page}{1}

\title{A study of fragmentation processes using a discrete element
    method}
\author{Ferenc~Kun and Hans~J.~Herrmann}

\maketitle
\centerline{ Laboratoire de Physique M\'ecanique des Milieux
            H\'et\'erog\`enes, C.N.R.S., U.R.A. 857}
\centerline{\'Ecole Sup\'erieure de Physique et Chimie
  Industrielle}
\centerline{10 rue Vauquelin, 75231 Paris, Cedex 05, France}

\vspace*{0.21truein}
\begin{abstract}
We present a model of solids made from polygonal cells connected
via beams. We calculate the macroscopic
elastic moduli from the beam and cell parameters.
This modellisation is particularly suited for the
simulation of fragmentation processes. We study the effects
of an explosion inside a circular disk and the impact of a
projectile and obtain the fragment size distribution.
We find that if breaking only happens under tensile forces
a layer on the free wall opposed to impact is first ejected.
In that case the distribution follows a power-law with
an exponent that in most cases is around two.
\end{abstract}

\section{Introduction}
\vspace*{-0.5pt}
\noindent
Fragmentation plays an important role in a wide variety of physical
phenomena. Examples range from geophysics to
astrophysics: fragments from weathering, coal heaps, rock fragments
from chemical and nuclear explosions, projectile collisions,
asteroids etc. Most of the measured fragment size distributions show
power law
behavior with exponents between 1.9 and 2.6 concentrating around 2.4
\cite{exp_tur} - \cite{exp_colli3}.
Power law behavior for small
fragment masses seems to be a common characteristic of the breaking of
brittle objects.

Fragmentation in one dimension was studied using thin glass rods
dropp\-ed vertically onto the floor \cite{exp_glass}.
Depending on the height from which the
rod was dropped the fragment size distribution varies from log -
normal to power law with increasing height.

Recently, Oddershede, Dimon and Bohr reported that the mass
distribution of fragments produced by impact experiments shows
universal power law behavior. The scaling exponents depend on the
overall morphology of the objects but are independent of the type of
the material. (Their experiment was performed with several different
kinds of materials, gypsum, soap, paraffin, potato.) Because of the
universal power law behavior found without any control parameter, they
concluded that the fragmentation process is a self - organized
critical process with exponent $\sim 1.6$ \cite{exp_bohr}.

Several theoretical approaches were established to describe
fragmentation. In one dimensional stochastic models
power law, exponential and log-normal fragment mass
distributions can be obtained depending
on the fracture point
distributions \cite{theo_glass}.

Discrete stochastic processes have also been studied as models for
fragmentation using cellular automata. In Ref.~8  two- and
three - dimensional cellular automata were proposed to model power
law distribution in shear experiments on a layer of uniformly sized
fragments.

In Ref.~9  an iterative stochastic process is studied as a
model of two- and three - dimensional discrete fragmentation.
Depending on the parameters of the model, log - normal and power law
distribution were found for the fragment size distribution.

The mean - field approach describes the time evolution of the
concentration $c(x,t)$ of fragments having mass $x$ through a linear
integro - differential equation \cite{redner}:
\begin{eqnarray}
  \label{mean}
  \frac{\partial c(x,t)}{\partial t} = -a(x) c(x,t) +
\int_x^{\infty} c(y,t)a(y)f(x|y)dy
\end{eqnarray}
where $a(x)$ is the overall rate at which $x$ breaks in a time
interval $dt$, while $f(x|y)$ is the relative rate at which $x$ is
produced from the break-up of $y$. With some further assumptions on
$f(x|y)$ exact results can be obtained but in physically interesting
situations the solution is
very difficult \cite{redner}.

Three - dimensional impact fracture processes of random materials were
modeled based on competitive growth of cracks \cite{soc}.
A universal power law
fragment mass distribution was found consistent with
self - organized criticality with an
exponent of $\frac{5}{3}$.

Recently, a two dimensional
dynamic simulation of solid fracture was performed
using a cellular model material \cite{potapov1,potapov2}.
The compressive failure of a
rectangular sample, the four - point shear failure of a beam and the
impact of particles with a plate and with other particles were
studied.

In this paper we present a two dimensional dynamic simulation of the
fragmentation of granular materials. We  establish a
two - dimensional model
for a  deformable, breakable, granular solid by connecting
unbreakable, undeformable  elements with elastic beams, similar as in
Refs.~12,13 .
The contacts between the
particles can be
broken according to a certain breaking rule, which takes into account
the stretching and bending of the connections.
The breaking rule contains two parameters to describe the
relative importance of the two breaking modes.
We address the question how the fragment mass distribution
depends on the breaking parameters.
After measuring the macroscopic elastic behavior of our granular solid
the model is applied to study
fragmentation caused by an explosion inside
a solid and by a projectile impact.

\section{The model}
\noindent
In order to study the fragmentation of granular solids we use Molecular
Dynamics (MD) simulations in two dimensions. A general overview of
MD simulations applied
to the field of granular materials can be found in
Ref.~14.  This
method calculates the motion of particles by solving Newton's
equations. In our simulation this is done using a Predictor Corrector
scheme.

Our model of a deformable, breakable granular solid is a generalization
of the model which was used earlier to study the flow of granular
materials \cite{hjt}. The model construction is composed of three major
steps, namely, the implementation of the granular structure of the solid,
the determination of the elastic behavior by the contact
force and the beam
models, and finally the breaking of the solid.
In this section, the three steps of the model construction are presented
in detail. \\

{\parindent 0cm
({\bf a}) Granularity
\vspace* {0.5cm}

In order to take into account the complex structure
of the granular solid we use arbitrarily shaped convex polygons.
To get an initial configuration of these polygons we construct a
vectorizable random lattice,
which is a Voronoi construction with a regularization procedure
 (see Ref.~16).
The advantage of the vectorizable random lattice compared to the
ordinary Poissonian Voronoi tessellation is that the
number of neighbors of each polygon is limited which makes the
computer code faster and allows to simulate larger systems.
The convex
polygons of this Voronoi  construction are supposed to model
the grains of the material, see Ref.~15.
In this way the structure of the
solid is built on a mesoscopic scale.
Each element is thought of as a large collection of atoms.
In our simulation however these
polygons are the smallest particles interacting
elastically with each other.
All the polygons have three continuous
degrees of freedom in two dimensions: the two coordinates
of the positions of the center of mass and the rotation angle.
}
\vspace* {0.5cm}

{\parindent 0cm

({\bf b}) Elastic behavior of the solid
}
\vspace* {1cm}

 The elastic behavior of the solid is captured in the following way:
The polygons are considered to be rigid bodies.
They are not breakable
and not deformable.
But they can overlap when they are pressed against each other
representing to
some extent the local deformation of the grains.
Usually the overlapping polygons have two intersection points.
These points define the contact line (see Fig. \ref{fig:overlap}).

In order to simulate the elastic contact force between touching grains
we introduce a
repulsive  force between the overlapping polygons. This
force is proportional to the overlapping area $A$ divided by a
characteristic length $L_c$ of the interacting polygon pair. Our
choice of $L_c$ is given by $1/L_c = 1/2(1/r_i
+1/r_{j})$, where $r_i, r_j$ are the diameters of circles of the
same area as the polygons.
This normalization is necessary in order to reflect
the fact that the spring constant is proportional to the elastic modulus
divided by a characteristic length. (In the case of a linear spring
this characteristic length is simply the equilibrium length of
the spring.)
The direction of the force is chosen to be
perpendicular to the contact line of the polygons.

The contact force $\vec{F}_{ij}$ between two particles is given by
\begin{eqnarray}
  \label{force}
 \vec{F}_{ij} = -\frac{YA}{L_c} \vec{n}
\end{eqnarray}
where $\vec{n}$ is the unit vector perpendicular to the contact line
(see Fig. \ref{fig:overlap})
and  $Y$ is the grain bulk Young modulus.
The friction of  the touching polygons
can be implemented according to
Coulomb's friction law (see also Ref.~15).  However it
turned out from the simulations that friction
does not play an important role in the fragmentation of
the granular solids.
Therefore, in the present simulations the friction term was omitted.

In order to keep the solid together it is necessary
to introduce a cohesion force  between
neighboring polygons. For this purpose we introduce beams,
which were extensively used recently
in crack growth models \cite{hans,beam}.
The centers of mass of neighboring
polygons are connected by elastic beams, which
exert an attractive, restoring force between the grains,
and can break in order to model
the fragmentation of the solid.

Because of the randomness contained in the Voronoi - tessellation
the lattice of beams is also random.
An example of a random lattice of beams coupled to the Voronoi polygons
can be seen in Fig. \ref{fig:randlatt}.

A beam between sites $i$ and $j$ is thought of as having a certain cross
section $S^{ij}$ giving to it not only
longitudinal but also shear elasticity. This cross section is the length
of the common side of the neighboring polygons in the initial
configuration. The length of the beam $l^{ij}$ is defined by the distance of
the centers of mass. The elastic behavior of the beams is governed by
two material dependent constants. For the beam between sites $i$ and $j$:
\begin{eqnarray}
  \label{const}
  a^{ij} &=& \frac{l^{ij}}{ES^{ij}} \\
  b^{ij} &=& \frac{l^{ij}}{GS^{ij}} \\
  c^{ij} &=& \frac{l^{ij^3}}{EI^{ij}}
\end{eqnarray}
where $E$ and $G$ are the Young and shear moduli of the beam,
$S^{ij}$ is the area of
the beam section, and $I^{ij}$ is the moment of inertia of the beam
for flexion.
A fixed value of $E$ was used for all the beams and
$b^{ij}$  was chosen to be $b^{ij} = 2a^{ij}$.
The length,
the cross section  and the moment of inertia of each beam
are determined by the random initial configuration
of the polygons as explained above.
The beam Young modulus $E$ and the grain bulk Young
modulus $Y$ are, in principal, independent.

In the local frame of the beam three continuous degrees of freedom are
assigned to both lattice sites (centers of mass) connected by the
beam, which are
for site $i$,
the two components of
the displacement vector  $(u_x^i,u_y^i)$ and a bending angle
$\Theta^i$.

For the beam between sites $i$ and $j$ one has the longitudinal force
acting at site $i$:
\begin{eqnarray}
  \label{longit}
  F_x^i &=& \alpha^{ij} (u_x^j - u_x^i) ,
\end{eqnarray}
the shear force
\begin{eqnarray}
  \label{shear}
 F_y^i  &=& \beta^{ij} (u_y^j - u_y^i) - \frac{\beta^{ij} l^{ij}}{2}
( \Theta^i + \Theta^j) ,
\end{eqnarray}
and the flexural torque at site $i$
\begin{eqnarray}
  \label{torque}
M^i_z &=& \frac{\beta^{ij} l^{ij}}{2} ( u_y^j -u_y^i +l^{ij}
\Theta^j)+\delta^{ij} l^{ij^2}(\Theta^j - \Theta^i)
\end{eqnarray}
where $\alpha ^{ij}= 1/a^{ij}$, $\beta^{ij} = 1/(b^{ij}+1/12c^{ij})$,
and $\delta^{ij} = \beta^{ij}(b^{ij}/c^{ij}+1/3)$ .
It can be shown that this beam model is a  discretisation of
the simplified Cosserat - equations of continuum
elasticity which should be
used to describe the elastic behavior of the granular solids, instead
of the Lam\'e  equations \cite{stephan}.
\vspace* {0.5cm}

{ \parindent 0cm
( {\bf c }) Breaking of the solid
}
\vspace* {0.5cm}

To model fragmentation it is necessary to
complete
the
model with a breaking rule, according to which the over-stressed beams
break.

For not too fast deformations  the breaking of a beam is only caused
by stretching and bending.
We  impose a breaking rule which takes into account
these two breaking modes, and which can reflect the fact
that the longer and thinner
beams are easier to break. We used a
breaking rule of the form of the von Mises plasticity criterion
\cite{hans}:
  \begin{eqnarray}
 \left( \frac{\epsilon}{t_{\epsilon}} \right)^2 + \frac{max
   (|\Theta^1|,|\Theta^2|)}{t_{\Theta}} &\geq 1 & \label{eq:mises}
\end{eqnarray}
where $ \epsilon = \Delta l /l$ is the longitudinal
strain of the beam, $\Theta^1$ and
$\Theta^2$ are the rotation angles at the two ends of the beam and
$t_{\epsilon}$ and $t_{\Theta}$ are threshold values for the
two breaking modes.
In the simulations we used the same threshold values $t_{\epsilon}$ and
$t_{\Theta}$
 for
all the beams.

Two sets of simulations were performed applying breaking
criteria of the type of Eq. (\ref{eq:mises}) but once only for stretched
beams, i.e. $\epsilon > 0$ and
once for stretched and compressed beams , i.e. for all $\epsilon$.
The first case is  physically more relevant since
it reflects the fact
that it is much harder to break a solid under compression than under
elongation.

The first term of Eq. (\ref{eq:mises}) takes into account the role of
stretching and the second term the role of bending.
Varying the threshold values  the relative importance of
the two modes in the beam  breaking can be changed.

During the simulation the left hand side of
Eq. (\ref{eq:mises}) is
evaluated at each iteration time step for all the existing beams,
which fulfill the strain conditions.
The breaking of beams means that those beams for which the condition
of Eq. (\ref{eq:mises}) holds
are removed from the calculation, i.e. their elastic constants
are set to zero.  Removed beams are never restored
during the simulation.

The surface of the grains, on which beams are broken represents cracks.
The energy of the broken beams is released
in creating these new crack surfaces inside the solid.

\section{Results}
\label{sec:results}
\noindent
 With the model described
above it is possible to perform a variety of  experiments.

First of all it is necessary to study the global elastic behavior of
our two dimensional solid assembled from cells.
Simulations were performed to measure the Young modulus and the
Poisson ratio of a rectangular sample.

In the present paper we are mainly interested in
fragmentation. We use the model to examine the
fragmentation of a heterogeneous
solid in two different
experimental situations,
namely,
the fragmentation of a disc-shaped solid caused by an explosion,
and the fragmentation of a rectangular
solid block due to the impact of  a high velocity projectile.

\subsection{Elastic properties}
\label{sec:elastic}
\noindent
In Ref.~12 it was shown that
the macroscopic elastic behavior of two dimensional materials
composed of particles with linear contacts
can be
characterized by two elastic constants, the Young modulus $K$ and the
Poisson ratio $\nu$, similar to the case of homogeneous isotropic solids.
It seems clear from the definition of the model that
these elastic constants depend on the properties of the
constituent particles, i.e.
on the shape of the grains, on the stiffness of the grain -
grain contacts (in our context the grain bulk Young modulus
$Y$ and the beam Young modulus $E$) and on the typical grain size.

For real materials the Young modulus $K$ and
the Poisson ratio $ \nu $ are usually determined by uniaxial loading
of the body.
In order to measure numerically the  elastic properties of our simulated
material we apply uniaxial loading on a two dimensional
rectangular sample, which
has linear extensions $L$ in the direction
of the loading and $S$ in the perpendicular direction. The
corresponding changes of the extensions due to loading are
denoted by $\Delta L$ and $\Delta S$.
During the simulations plane -
stress conditions are used, i.e. it is assumed that there is no
stress in the direction out of the plane of the material. The
boundaries parallel to the loading are free in the simulation.

In the case of uniaxial loading of a rectangular sample  under
plane - stress
conditions, the Young modulus $K$ is the ratio between the external
force $F$
per unit length
acting on the solid and the strain in the
direction of the loading, $F / S$ = K $\Delta L / L$.
The Poisson ratio $\nu$ is the ratio between
the strain in the direction perpendicular to  the loading
and the strain in the direction of the loading, $\Delta S/S = - \nu
\Delta L / L$.

To avoid the disturbing effect of the elastic waves induced by the
loading, the numerical experiment is performed in the following way
(see also Ref.~12):
The two opposite boundaries of the solid start to move with zero
initial velocity and non-zero
acceleration. When a certain velocity is reached the acceleration is
set to zero and the velocity of the boundaries is kept fixed.
With this slow loading the vibrations of the solid can be reduced
drastically compared to the case when the boundaries start to move with
non - zero initial velocity. A further way to suppress artificial vibrations
is to introduce a small dissipation (friction or damping) between the grains.
This dissipation has to be
small enough not to affect the quasi-static results.
To obtain the values of $K$ and $ \nu $, the force
acting on the boundary layer, and the horizontal
and vertical extensions of the sample were monitored during the loading.

Fig. \ref{fig:young_meas} shows an example of the results of the
measurement of the Young
modulus, the horizontal stress $\sigma = F/S$ as a
function of the vertical strain $\epsilon = \Delta L/L$.
The values of $K$ were extracted from the slope of the straight lines.
It can be
seen  that  elongation and compression of the system are not
symmetric. The solid is more stiff under compression.
The reason of the asymmetry is that in the case of elongation
practically only the beams act but under compression one measures the
common effect of the beams and the overlap force giving rise to a
larger effective Young modulus.

Calculations were performed for elongation and compression of
the sample fixing the value of the grain bulk Young modulus $Y$ and
varying the beam Young modulus $E$. The results for $K$ are
shown in Fig. \ref{fig:young}.
In the case of elongation $K$ is a linear function of $E$ as
expected.
For large $E$
the elongation and compression curves are parallel. The
difference between them is determined by the grain bulk Young modulus
$Y$ and by the geometry of the
Voronoi tessellation.
Small values of $E$ compared to $Y$ means that the cohesive
force in the solid is small with respect to the repulsive overlap force.
Under compression the polygons
can move perpendicular to the loading in the direction of the free
boundaries in the limiting case of small cohesion.
Because of the dense packing
(the initial packing fraction is one) the deformation
tends to increase the overall volume resulting in dilatancy
 \cite{bashir}.
This effect gives rise to a small effective Young modulus,
such that $K$ is
even smaller than the grain bulk Young modulus $Y$.

For the Poisson ratio $\nu$ a similar asymmetry
of elongation and
compression is observed
as shown in Fig. \ref{fig:poisson}. The values
of $\nu$ for elongation are smaller than for compression but
for large $E$  both values approach the Poisson ratio of the
pure beam lattice.
In the
limiting case of small cohesion $\nu$ can even exceed unity,
due to the dilatancy mentioned above.

\subsection{Explosion of a disc - shaped solid}
\label{sec:expl}
\noindent
The catastrophic fragmentation of solids will be studied in
two different experimental situations: through an explosion which takes
place inside the solid and through the impact with a projectile
(stroke with a
hammer).
This chapter is devoted to study the explosion.

In the explosion experiment
the detonation
takes place in the center of a solid disc.
The granular solid with disc - like shape was obtained starting
from the Voronoi tessellation of a square and cutting out a circular
disc in the center, see Fig. 2.

In the center of the solid we choose one polygon, which plays the role
of the explosive. Initial velocities are given to the neighboring
polygons perpendicular to their common sides
with the
central one. The sum of the  initial linear momenta
has to be zero, reflecting the spherical symmetry of the
explosion. From these two constraints it follows that for a polygon
having mass $m$ and a common boundary of length $S$ with the explosive
center, the initial velocity is proportional to $\frac{S}{m}$. The sum
of the initial
kinetic energies defines the energy $E_o$ of
the explosion. (For the parameter values and the initial conditions
of the simulation see Table \ref{table_0}.)

As a result of these initial conditions a circularly symmetric
outgoing compression wave is generated in the solid.
In our context this means that there is a well -  defined
shell where the average longitudinal
strain of the beams $< \epsilon>  = <\Delta l / l>$ is negative.
This compression wave is not homogeneous in the sense that not all the
beams in this region are compressed . If the angle of a beam with
respect
to the radial direction is close to $\pi / 2$ a beam can be slightly
elongated
within the compression wave.

Since the overall shape of the solid has the same symmetry as the
compression
wave it is possible to avoid
geometrical asymmetries, which would arise for example in the
explosion of a
rectangular sample due to the corners.
In the disordered solid the initial
compression wave gives rise to a complicated stress distribution,
in which the over-stressed beams break according
to the breaking rule Eq. (\ref{eq:mises}).
The simulation is stopped if there is no beam breaking
during 300 successive time steps. Free boundary conditions were used in
all simulations.

Due to the beam breaking the solid eventually breaks apart, i.e. at
the end of the process it consists of well separated groups of
polygons. These groups of polygons,  connected by the remaining beams, are
the fragments.
In the simulation of the explosion we are mainly interested in the
time evolution of the fragmentation process and the
mass distribution of fragments at the end of the process
as a function of the breaking parameters.

All the calculations were performed on the CM5 of the CNCPST in Paris.
We used the farming method, i.e. the same program runs on a variety of
nodes with different initial setups. In our case 32 nodes were used
with different seeds for the Voronoi generator, i.e. with differently
shaped Voronoi cells.

\subsubsection{The time evolution of the explosion}
\noindent
We performed two sets of calculations one  allowing the beam breaking
solely
under elongation and another under elongation and compression,
keeping all the other parameters of the simulation fixed.
The evolution of the explosion and the resulting breaking
scenarios  are different in these two cases.

We can distinguish
two regimes in the time evolution of the explosion:
The initial regime is controlled
by the compression wave and the disorder of the solid.
The amplitude of the shock wave is proportional to the ratio of
the average initial velocity of the polygons to the longitudinal
sound speed of the solid. The width and the speed of the
wave are mainly determined by the grain size and the  Young moduli.

If the beams are not allowed to break under compression the
compression wave can go outward
almost unperturbed and an elongation
wave is formed behind it. The elongated beams break according to the
breaking rule Eq. (\ref{eq:mises}) only for $\epsilon > 0$.
Due to the elongation wave  a
highly damaged region is created in the vicinity of the explosive
center
where
practically all the beams are broken and all the fragments are single
grains.
This highly damaged region is called the mirror spot.
Since the breaking of the beams, i.e. the formation of cracks in the solid,
dissipates energy  after some time the growth of the damage stops.
The size of
this mirror spot is determined by the initial energy of the explosion,
by the dissipation rate and by the
breaking thresholds (see Fig. \ref{fig:exp_elong}).

During and after the formation of the mirror spot when the outgoing
compression and elongation waves go through the solid the weakest
(i.e. the longest and thinnest) beams
break in an uncorrelated fashion
creating isolated cracks in the system. The uncorrelated beam breaking
is dominated by the
quenched disorder of the solid structure.
This first uncorrelated regime of the explosion process
lasts till the compression wave reaches
the free boundary of the solid.

{}From the free boundary the compression wave is reflected back with
opposite phase generating
an incoming elongation wave.
The
constructive interference of the incoming and outgoing
elongation waves gives rise
to  a
highly stretched zone close to the boundary.
The beams having small angle with respect to the radial direction have
the largest elongation.
In this zone a
large number of beams break  causing usually the complete break-off of a
boundary layer along the surface of the solid.
The thickness of this detached layer is roughly half the width of the
incoming elongation wave (see Fig.\ref{fig:exp_elong}).
The fragments of this boundary layer fly away in the radial direction with
a high velocity carrying with them a large portion of the total energy in the
form of their kinetic energy.
After that the system starts to expand. This overall expansion
initiates cracks going from inside to outside and from
outside to inside.
The branching of single cracks and the interaction of different
cracks give rise to the final fragmentation of the solid
(Fig.\ref{fig:exp_elong}).
This second
part of the evolution of the explosion process is dominated by the
correlation of the cracks.

The propagation of the elastic waves when the beam brea\-king is
swit\-ched off is
pre\-sen\-ted in Fig. \ref{fig:wave}.
One can observe the peak of the initially imposed shock,
the propagation of the
compression and elongation waves and the formation of the highly
stretched zones at the boundary.

If the beams are allowed to break  also under compression the time
evolution of the process is significantly different.
In the vicinity of the explosion center the initial compression wave
governs the formation of the mirror spot. Where the compression wave
passes all the beams break
as long as the amplitude of the compression wave is
larger than the  threshold strain of the breaking criterion
Eq. (\ref{eq:mises}).
Thus when the energy dissipation stops the growth of the damage the
amplitude of the remaining compression wave is weaker and the size of
the mirror spot is larger than it was in the previous case (Fig.
\ref{fig:exp_comp}). A
significant part of the initial energy of the explosion is dissipated
by the beam breaking and is transfered to the kinetic energy of the single
polygons. So, less fragmentation can occur in the outer regions of the
disc as compared to the case where breaking only occurs under elongation.

The compression wave reflects back from the free boundary but the
constructive interference does not cause substantial damage along the surface
(Fig.\ref{fig:exp_comp}).
That is why the final breaking picture is different from the previous case.

\subsubsection{The fragment mass distribution}
\noindent
Fig. \ref{fig:exp_elong} and Fig. \ref{fig:exp_comp} show that
the final breaking scenarios for the two breaking criteria
are significantly
different.
When the beams are also allowed to break under compression
there are only a few small fragments (apart
from the single polygons) and a few much larger ones.
This results
in a fast decay of the fragment size
distribution.
The resulting fragment mass histogram is presented in
Fig. \ref{fig:mind}.
$F(m)$ denotes the number of
fragments with mass $m$ divided by the total number of fragments.
As can be seen from the logarithmic plot in Fig. \ref{fig:mind} a,
an exponentially decaying function can be
a reasonable fit to these results for approximately one order of magnitude in
mass.
The quality of the fit is  demonstrated in
Fig. \ref{fig:mind} b using a semi-logarithmic plot.

The application of Eq. (\ref{eq:mises}) for $\epsilon > 0$ only,
is physically more
relevant. In the following we present
results of the mass distribution of the
fragments as a function of the breaking
parameters allowing the beam breaking solely under stretching.
Changing  $t_{\epsilon}$ and $t_{\Theta}$ in the
breaking criterion Eq. (\ref{eq:mises}) one can vary the relative
contribution of the stretching and bending breaking modes, which can
affect the crack formation and the final breaking scenario. The
question is whether the mass distribution of the fragments is
invariant under a change of $t_{\epsilon}$ and $t_{\Theta}$.

We performed simulations alternatively fixing $t_{\epsilon}$ and
$t_{\Theta}$
and changing the value of the other one, keeping all the other
parameters of the simulations fixed.
In both cases the fraction
of broken beams $p$ was monitored as a function of $s/l$, where $s$
and $l$ are the cross section and the length of a beam, respectively:
\begin{eqnarray}
  \label{fraction}
  p(s/l) = b \frac{N_{broken}(s/l)}{N(s/l)},
\end{eqnarray}
where  $N(s/l)$ and $N_{broken}(s/l)$ denote the number of beams in
the sample
and the number of broken beams having a ratio of $s/l$. $b$ is a
normalization constant.
Fig. \ref{fig:prob_elo} and Fig. \ref{fig:prob_ang}
show the results for fixed $t_{\Theta} = 4^o$
varying the threshold strain $t_{\epsilon}$ from $1\%$ up to $6\%$
and for fixed $t_{\epsilon}= 3\%$ varying the threshold angle
$t_{\Theta}$ from $1^o$ up to $7^o$, respectively.
The curves are normalized in such
a way that their integrals are equal to the overall fraction of broken
beams  $p_o$, i.e. $\int p(s/l) d(s/l) = p_o$. Here $p_o = N_{broken}/N$,
where $N$ and $N_{broken}$ denote the total number of beams in the sample
and the total
number of broken beams, respectively. One can observe that $p(s/l)$ is
a monotonically decreasing function for all parameter
values, which shows that longer and thinner beams are always
easier to break, as expected.
In Fig. \ref{fig:prob_elo} for small $t_{\epsilon}$ and in
Fig. \ref{fig:prob_ang} for small $t_{\Theta}$ the solid is more
fragile, so it undergoes strong damage. This results in a larger
$p_o$ and a rather flat breaking fraction $p(s/l)$. The curves belonging
to increasing $t_{\epsilon}$ and $t_{\Theta}$  lie below each
other.
Since by
increasing a breaking threshold,  the contribution of the
corresponding breaking
mode becomes
less important, the fixed  mode starts to dominate the
breaking and determines the limiting curve.
The results obtained by switching completely off one of the breaking
modes confirm this argument.

The fragment mass
histograms are presented in Fig \ref{fig:elo_dist} and
Fig. \ref{fig:bend_dist}.
The lower cutoff of the histograms is determined by the size of the
unbreakable polygons (smallest fragments) while the upper cutoff is
given by the
finite size of the system (largest fragment).
Larger values of the overall breaking fraction $p_o$ entails that the system
is broken into smaller pieces.

The histograms follow a power law for practically all the parameter
pairs for at least one order of magnitude in mass, such that we seem to
have
\begin{eqnarray}
\vspace{0.4cm}
  \label{power}
  F(m) \sim \alpha m^{-\beta} .
\vspace{0.4cm}
\end{eqnarray}
The effective exponents $\beta $ were obtained
from the estimated slopes of
the curves. The results are presented in Table \ref{table_1}.
Apart from the case
of extremely small breaking parameters the exponent $\beta$ only slightly
varies around $\beta = 2.0$, indicating a more or less universal
behavior within the accuracy, with which $\beta$ was determined ($\pm 0.05$).
Fig. \ref{fig:crossover} shows that increasing
$t_{\epsilon}$ and $t_{\Theta}$ a crossover appears in $F(m)$,
i.e. there are two exponents.
As we have mentioned, the
initial compression wave gives rise to the break-off of a boundary
layer, the thickness of which is determined by the wavelength and by the
breaking parameters. For larger
 $t_{\epsilon}$ and $t_{\Theta}$, when the
system is tougher, a detachment of the fragmentation of
the boundary layer and the bulk of the solid appears. The boundary gives the
main contribution to the small fragments (containing $\sim$ 2 -- 6
polygons) and the fragmentation of the bulk dominates
the range of the larger ones.
In these cases the exponents $\beta$ in Table \ref{table_1} were
obtained in the limit of large fragments.

\subsection{Impact of a projectile}
\label{sec:imp}
\noindent
Besides the explosion a catastrophic fragmentation of solids can also be
generated by an impact with a projectile.
In nature one of the most spectacular examples is that the size
distribution of meteorites and asteroids shows power law
behavior. These objects
are believed to have originated from the fracture of primitive planets
due to collisions \cite{exp_tur,exp_colli3}.

We applied our model to study the fragmentation of a rectangular solid
block (e.g. a block of concrete) due to an impact allowing the beam
breaking solely under stretching $\epsilon > 0$.
One polygon at the lower middle part of the block is given a high
velocity directed inside the block simulating an elastic collision
with a projectile. The boundary conditions and the stopping condition
were the same as in the explosion experiment. The breaking thresholds
were chosen $t_{\epsilon} = 3\%$ and $t_{\Theta}= 4^o$.

The evolution of the fragmenting solid block is presented in
Fig. \ref{fig:impact}. As in the case of the explosion,
the initially generated compression
wave plays a significant role.
Since the energy of the collision is concentrated around the impact
site of
the projectile the damage is the largest in that region. The
completely destroyed zone, where all the beams are broken
stretches inside the solid in the forward direction
resulting in the break-up of the solid. When
the shock wave reaches the boundary at the side of the solid
opposite to the collision point it gives rise to the break-off of a
boundary layer. The fragments of this layer fly away
in the forward direction with a high velocity.
Some small fragments from the vicinity of the collision
point are scattered backward.
The damage in the direction perpendicular to the projectile is not
strong, the broken boundary layer is thicker and the speed of the
fragments is smaller.

Results of laboratory experiments on high velocity impacts can be
found in Refs.~3,4. In Ref.~4
a picture series obtained by a high speed camera is presented showing
the time evolution of an impact experiment. Our results are
qualitatively in good agreement with the experimental observations.

The resulting fragment mass histogram $F(m)$ is presented in
Fig. \ref{fig:coll_mass}. Similarly to the explosion experiment, $F(m)$
shows power law behavior for approximately one order of magnitude in
mass. The value of the effective exponent is $\beta = 1.98 \pm 0.05$.
\\

\section{Conclusion}
\noindent
We have studied shock fragmentation in two dimensions using a cell
model similar to the one introduced by Potapov et al
\cite{potapov1,potapov2}. We found a qualitative difference in the
breaking process when breaking occurs under traction or under both,
traction and compression. If only traction force can produce cracks a
power-law distribution of the fragment sizes is evidenced with a
rather universal exponent around two in good agreement with most
experimental observations. Interesting is the effect of lamination in
which  a surface layer half the width of the elastic wave is detached
and ejected with high velocity. This effect has been observed
experimentally in high speed impacts.

 With the present study we have evidenced that the full treatment of
elasticity and wave propagation are necessary if one wants to
reproduce subtleties of fragmentation as lamination or the dependence on
compressive breaking. Still our study makes a certain number of
technical simplifications which might be important for a full
quantitative grasp of fragmentation phenomena. Most important seems to
us the restriction to two dimensions, which should be overcome in
future investigations. The existence of elementary, non-breakable
polygons restricts fragmentation on lower scales and hinders us from
observing the formation of powder of a shattering transition
\cite{redner}.

An advantage of our model with respect to most other fragmentation
models is that we can follow the trajectory of each fragment, which is
often of big practical importance and that we know how much energy
each fragment carries away. The polygonal structure of our solid
allows us to realistically model granular or polycrystalline matter,
and considering as well cell repulsion as beam connectivity gives us a
rich spectrum of possibilities ranging from breaking through bending to
the effect of dilatancy. If one or the other mechanism is turned off
we have the extreme cases of an elastic homogeneous solid and a
compact dry granular packing.

\section*{Acknowledgment}
\noindent
We are grateful to S. Roux, F. Tzschichholz and
S. Schwarzer for the helpful discussions.
We would like to thank Charles S. Campbell for sending us preprints
of his work.
F. Kun acknowledges the financial support
of the Hungarian Academy of Sciences.

\newpage

\begin{figure}[!h]
\vspace{0.3cm}
\begin{center}
\epsfig{bbllx=118,bblly=435,bburx=515,bbury=692,file=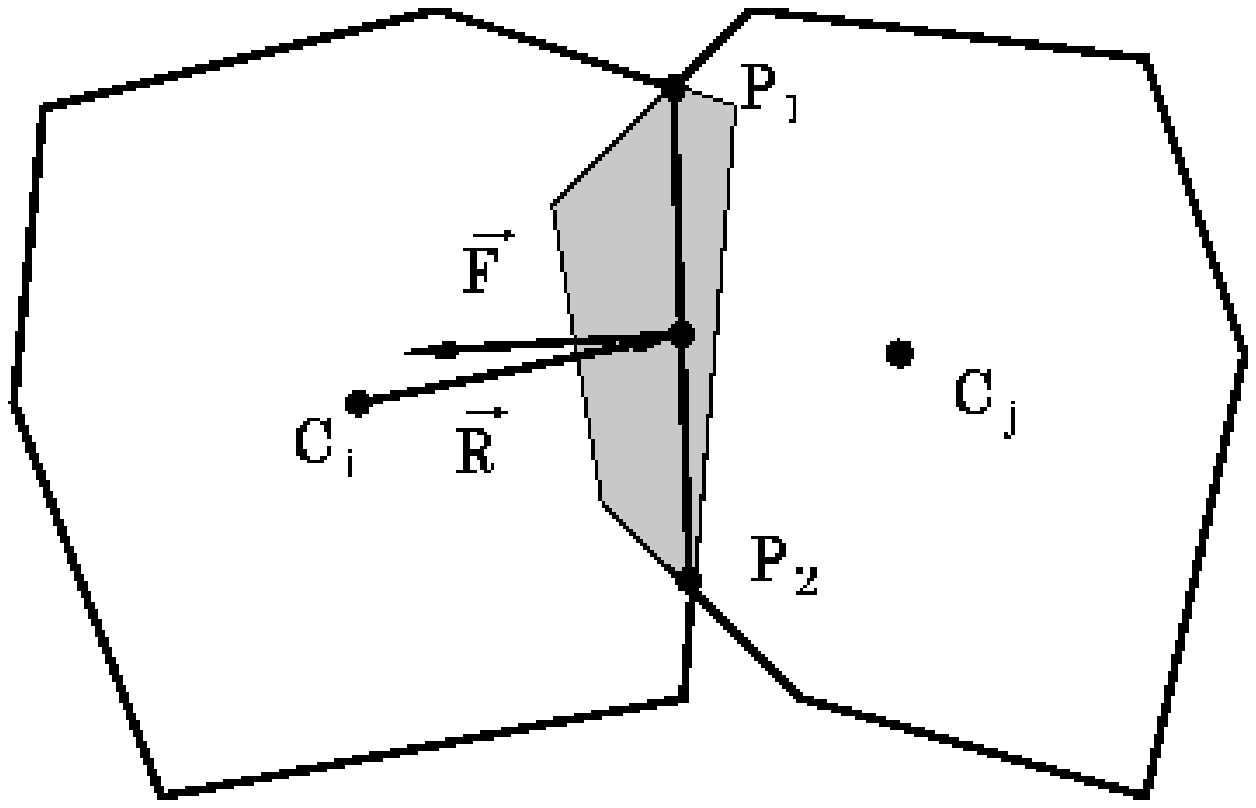,width=15cm}
\caption{To calculate the elastic contact force $\vec{F}$ between
  two particles
  one has to obtain the overlap area. The intersection points $P_1$,
  $P_2$ define the contact line $\vec{P_1P_2}$. The force is
  applied at the center of the contact line $\vec{R}$ and the direction of the
  force is perpendicular to $\vec{P_1P_2}$.}
\label{fig:overlap}
\end{center}
\end{figure}

\begin{figure}[!h]
\vspace{0.7cm}
\begin{center}
\epsfig{bbllx=170,bblly=330,bburx=480,bbury=620,file=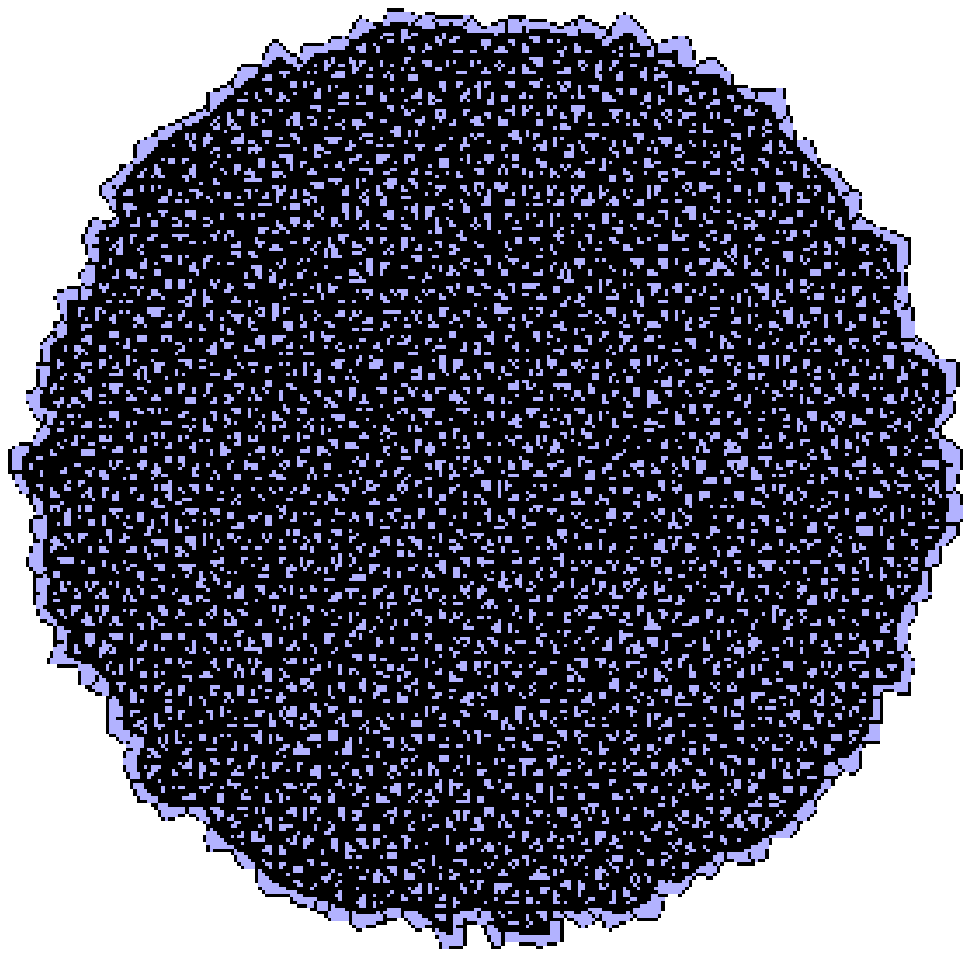,width=12cm}
\caption{Elastic beams connecting the Voronoi polygons in a disc -
  shaped sample. Because of the randomness in the Voronoi tessellation
  one has also a  random lattice of beams.}
\label{fig:randlatt}
\end{center}
\end{figure}

\begin{figure}[!h]
\vspace{0.5cm}
\begin{center}
\epsfig{bbllx=170,bblly=330,bburx=480,bbury=620,file=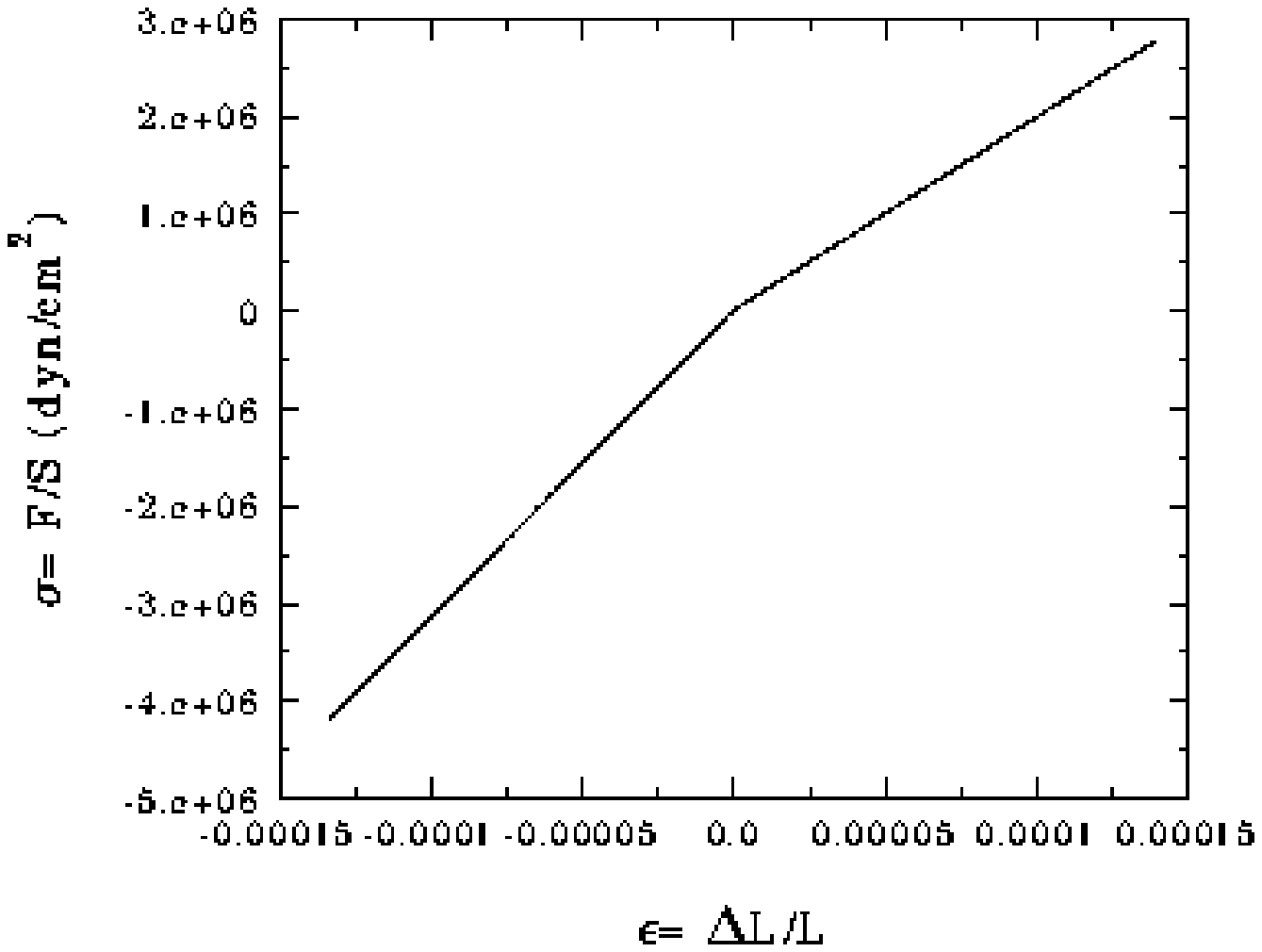,width=12cm}
\caption{Typical result of a measurement of the effective Young
  modulus $K$, the horizontal stress $\sigma =F/S$ as a function of
  the vertical strain $\epsilon = \Delta L/L$.
  Positive and negative strain means elongation and compression of the
  solid, respectively.
  The values of $K$ were extracted from the slopes of the straight
  lines. One can observe the asymmetry of elongation and compression.}
\label{fig:young_meas}
\end{center}
\end{figure}

\begin{figure}[!h]
\vspace{0.2cm}
\begin{center}
\epsfig{bbllx=170,bblly=330,bburx=480,bbury=620,file=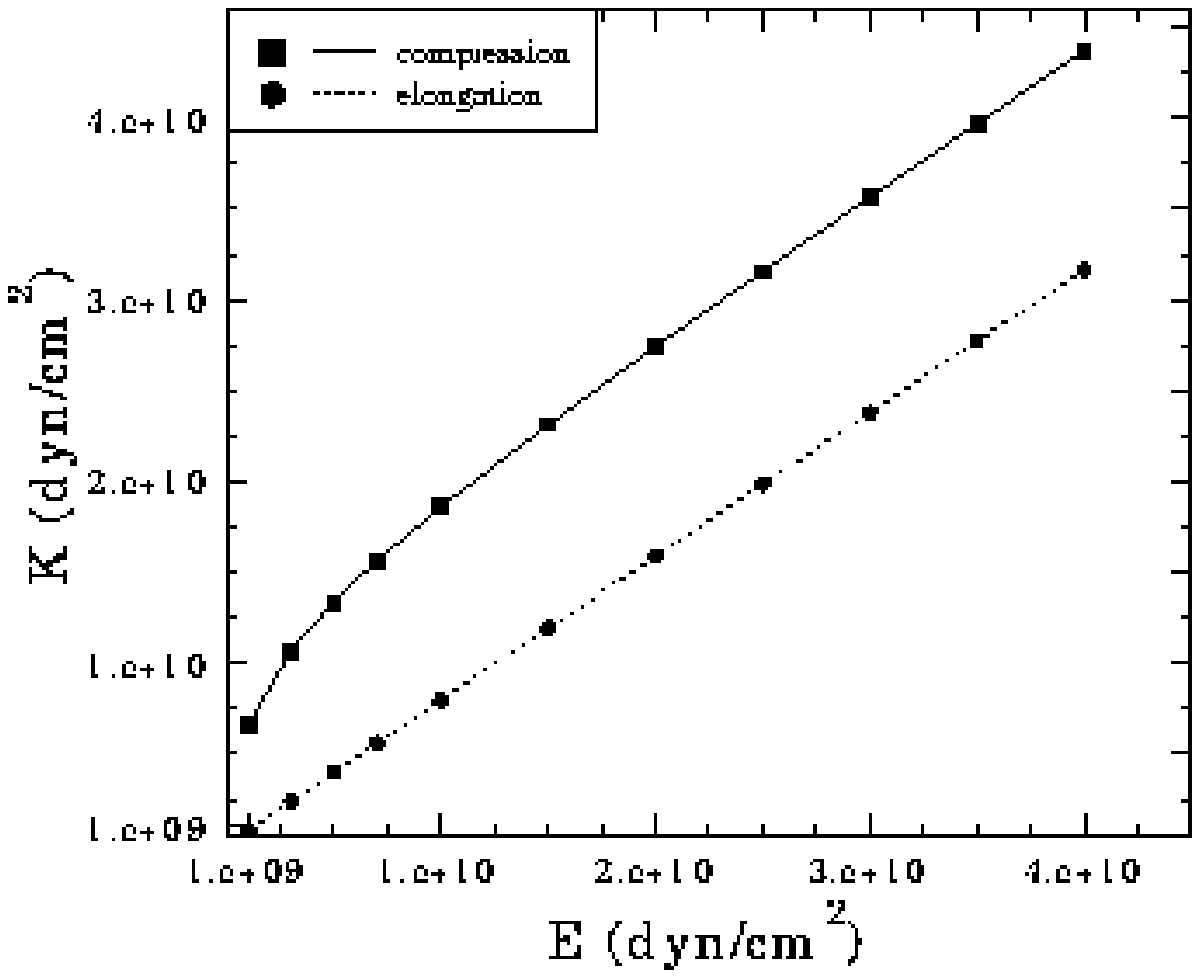,width=12cm}
\caption{The effective (macroscopic) Young modulus $K$ of the granular
  solid measured under compression and elongation varying the beam
  Young modulus $E$ between $10^9 dyn/cm^2$ and $4 \cdot 10^{10}
  dyn/cm^2$. The grain bulk Young modulus was fixed to be
  $Y = 10^{10} dyn/cm^2$.}
\label{fig:young}
\end{center}
\end{figure}

\begin{figure}[!h]
\vspace{0.2cm}
\begin{center}
\epsfig{bbllx=170,bblly=330,bburx=480,bbury=620,file=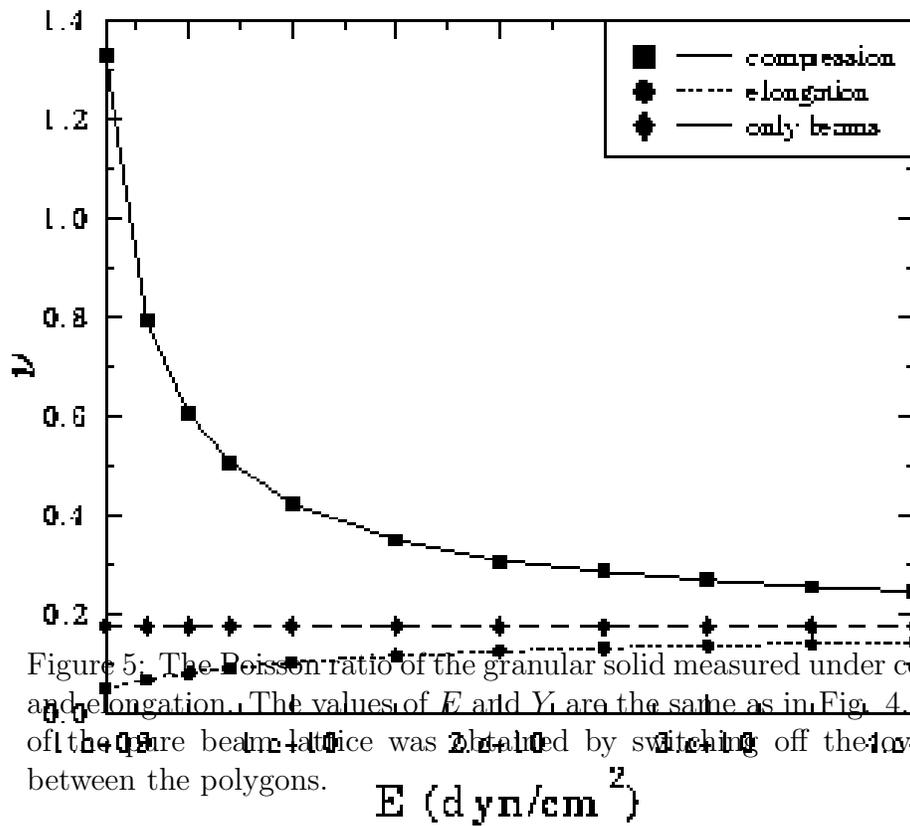,width=12cm}
\caption{The Poisson ratio of the granular solid measured under
  compression and elongation. The values of $E$ and $Y$ are the same
  as in Fig. \ref{fig:young}. The curve of the pure beam lattice was
  obtained by switching off the overlap force between the polygons.}
\label{fig:poisson}
\end{center}
\end{figure}

\begin{figure}[!h]
\vspace{0.0cm}
\begin{center}
\epsfig{bbllx=154,bblly=340,bburx=477,bbury=631,file=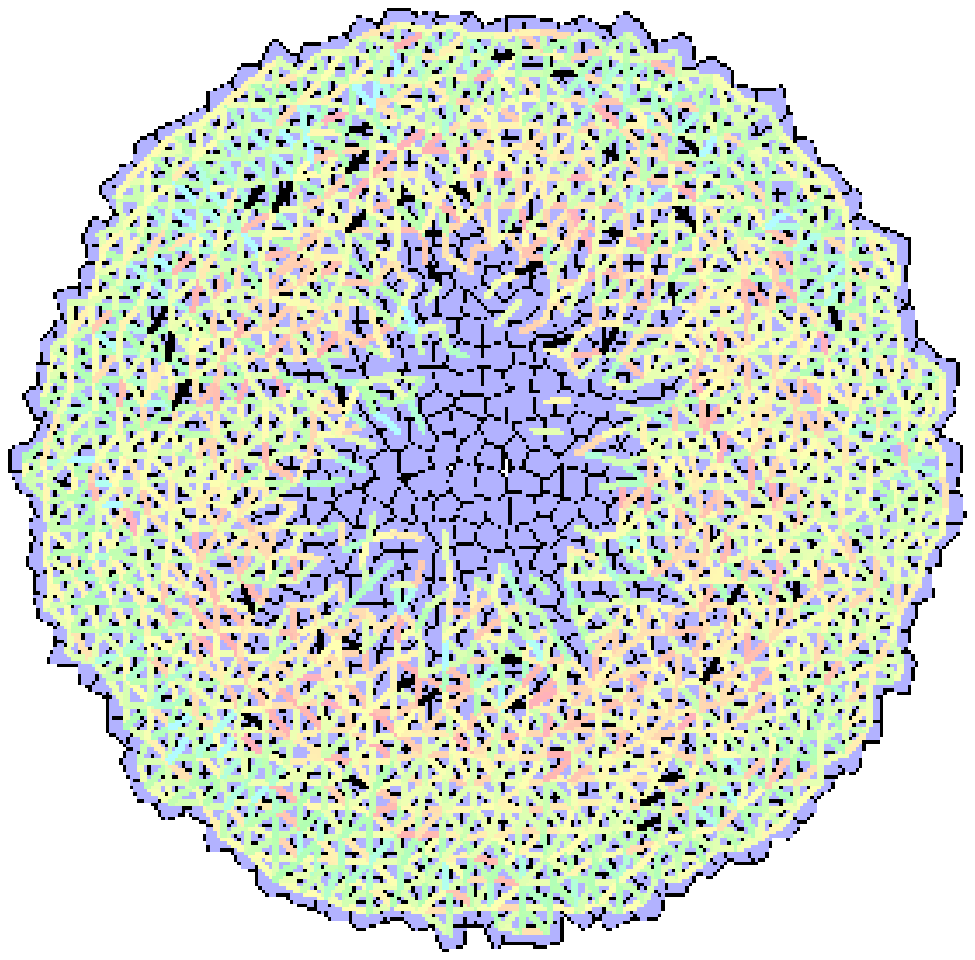,width=5.5cm}
\\
\epsfig{bbllx=231,bblly=162,bburx=481,bbury=400,file=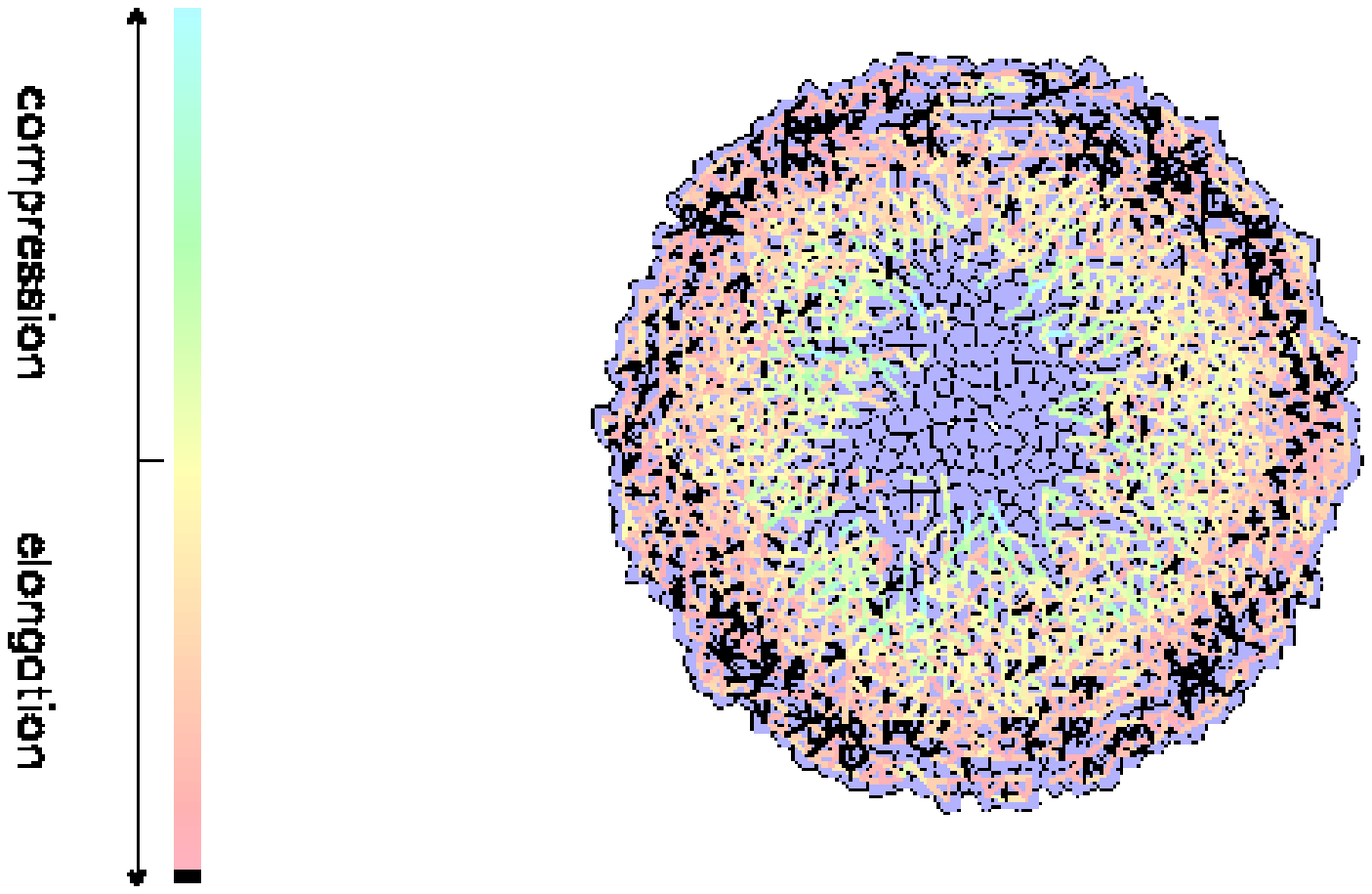,width=5.5cm}
\\
\epsfig{bbllx=170,bblly=325,bburx=480,bbury=670,file=bound3_1.ps,width=5.5cm}
\\
\caption{Explosion of a disc - shaped solid when only the stretched
  beams $(\epsilon > 0)$
  are allowed to break.
  Snapshots of the evolving system are presented when the
  initial compression wave reaches the boundary of the solid
  ($t=0.0001 s$),
  the constructive interference of the incoming and outgoing
  elongation waves breaks the boundary layer ($t=0.001 s$) and the
  final breaking scenario ($t=0.004 s$). The color scale shows the
  color code for the beams. Black indicates that the strain is close to the
  stretching threshold $t_{\epsilon}$.}
\label{fig:exp_elong}
\end{center}
\end{figure}
\begin{figure}[!h]
\vspace{0.0cm}
\begin{center}
\epsfig{bbllx=147,bblly=73,bburx=482,bbury=758,file=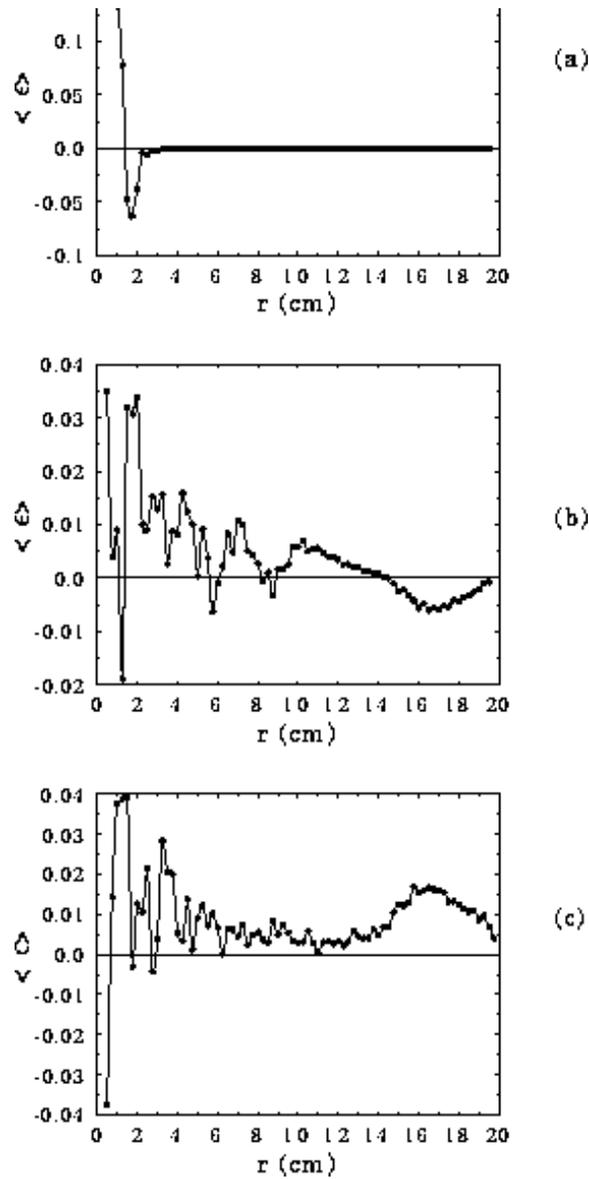,width=9cm}
\caption{Propagation of the elastic wave in the disc - shaped solid.
  $(a)$ after $t=10^{-5} sec$ of the initial hit, $(b)$ after $t= 3 \cdot
  10^{-4} sec$ the compression wave is
  approaching the boundary and  $(c)$ after $t=5 \cdot 10^{-4}$ the
  constructive interference of the
  incoming and the outgoing elongation waves.  }
\label{fig:wave}
\end{center}
\end{figure}

\begin{figure}[!h]
\vspace{0cm}
\begin{center}
\epsfig{bbllx=180,bblly=330,bburx=480,bbury=620,file=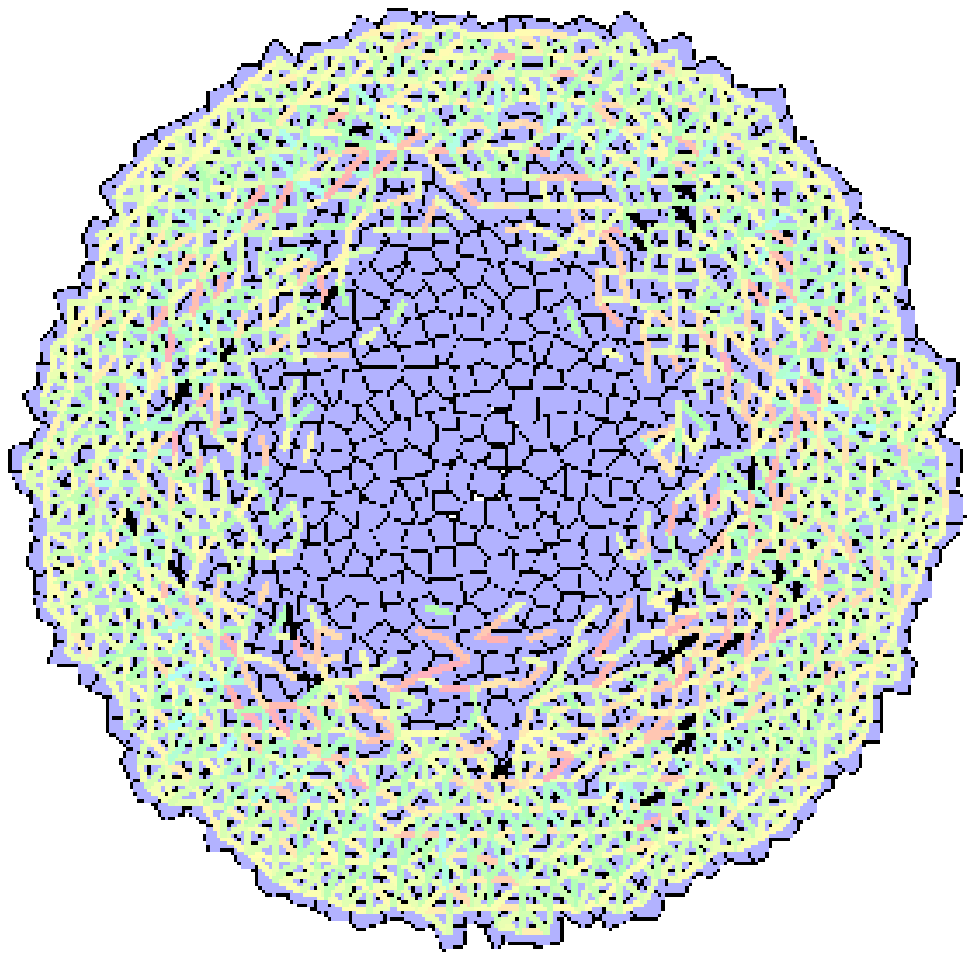,width=5.5cm}
\\
\epsfig{bbllx=271,bblly=322,bburx=438,bbury=480,file=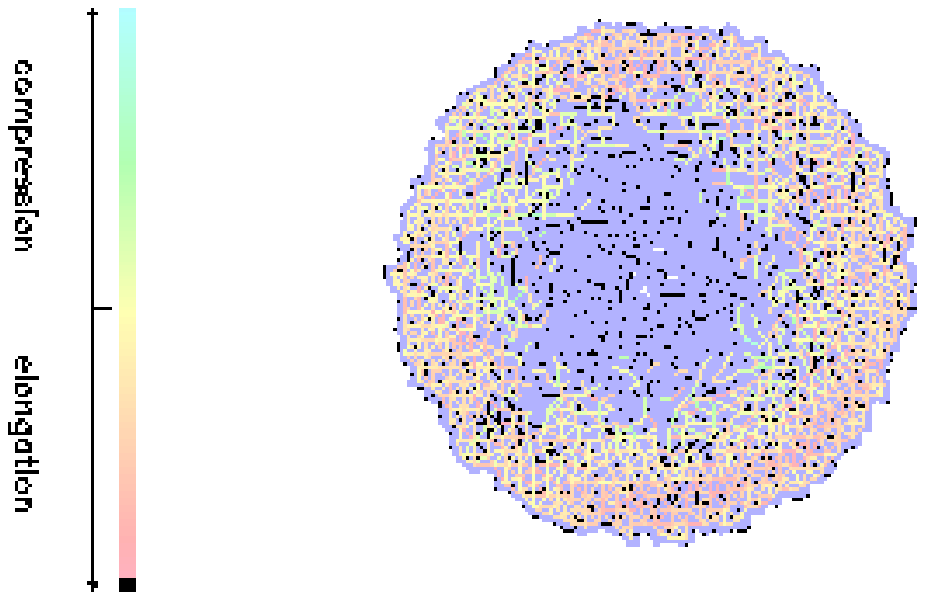,width=5.5cm}
\\
\epsfig{bbllx=160,bblly=325,bburx=499,bbury=667,file=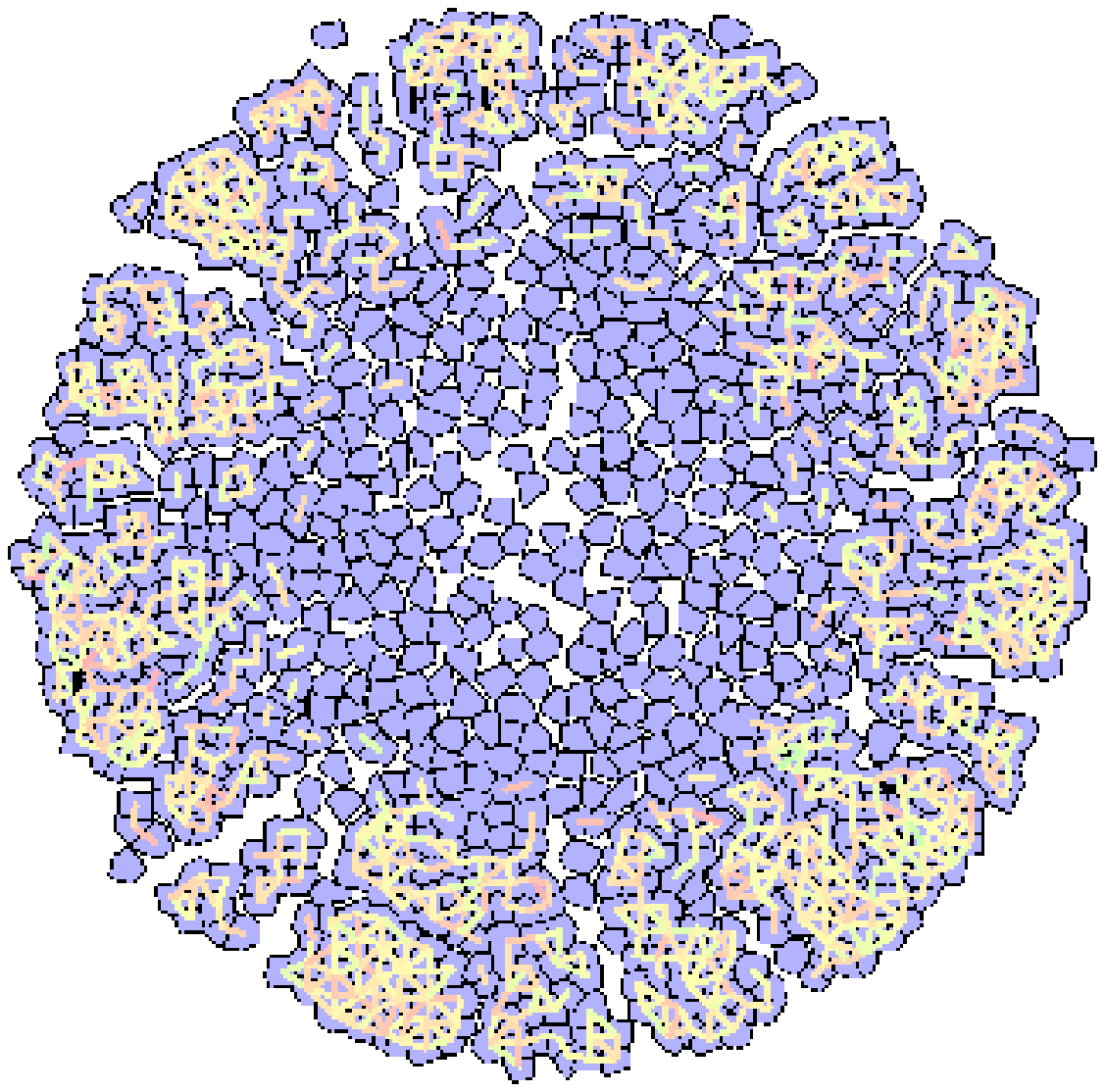,width=5.5cm}
\\
\caption{Explosion of a disc - shaped solid allowing the beam breaking
  under compression and elongation. The snapshots are taken at the
  same times as in Fig.\ref{fig:exp_elong}.}
\label{fig:exp_comp}
\end{center}
\end{figure}

\begin{figure}[!h]
\vspace{-0.1cm}
\begin{center}
\epsfig{bbllx=130,bblly=73,bburx=500,bbury=682,file=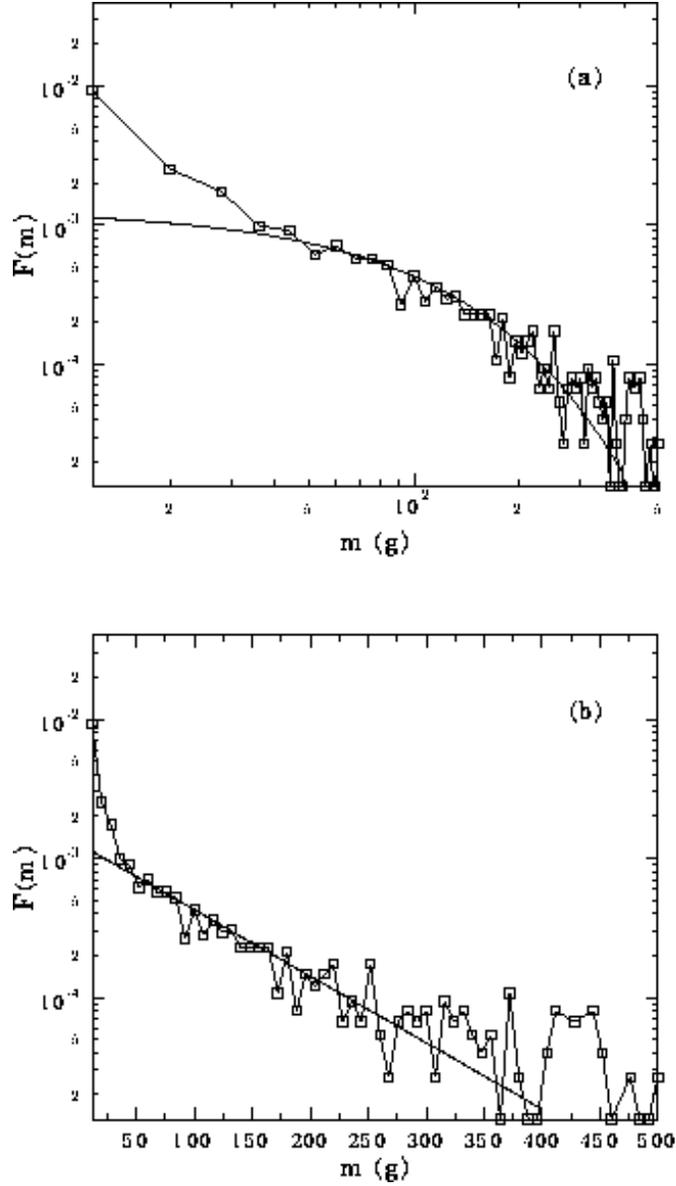,width=10cm}
\caption{Fragment mass histogram
  allowing the beam breaking under
  compression and elongation. The breaking thresholds were chosen
  $t_{\epsilon}= 3\%$ and $t_{\Theta} = 4^o$. (a) shows on a
  logarithmic plot that $F(m)$ does not exhibit power law behavior and an
  exponential function seems to be a reasonable fit (bold line).
  (b) demonstrates the quality of the fit on a semi -- logarithmic plot.}
\label{fig:mind}
\end{center}
\end{figure}

\begin{figure}[!h]
\begin{center}
\epsfig{bbllx=117,bblly=374,bburx=501,bbury=686,file=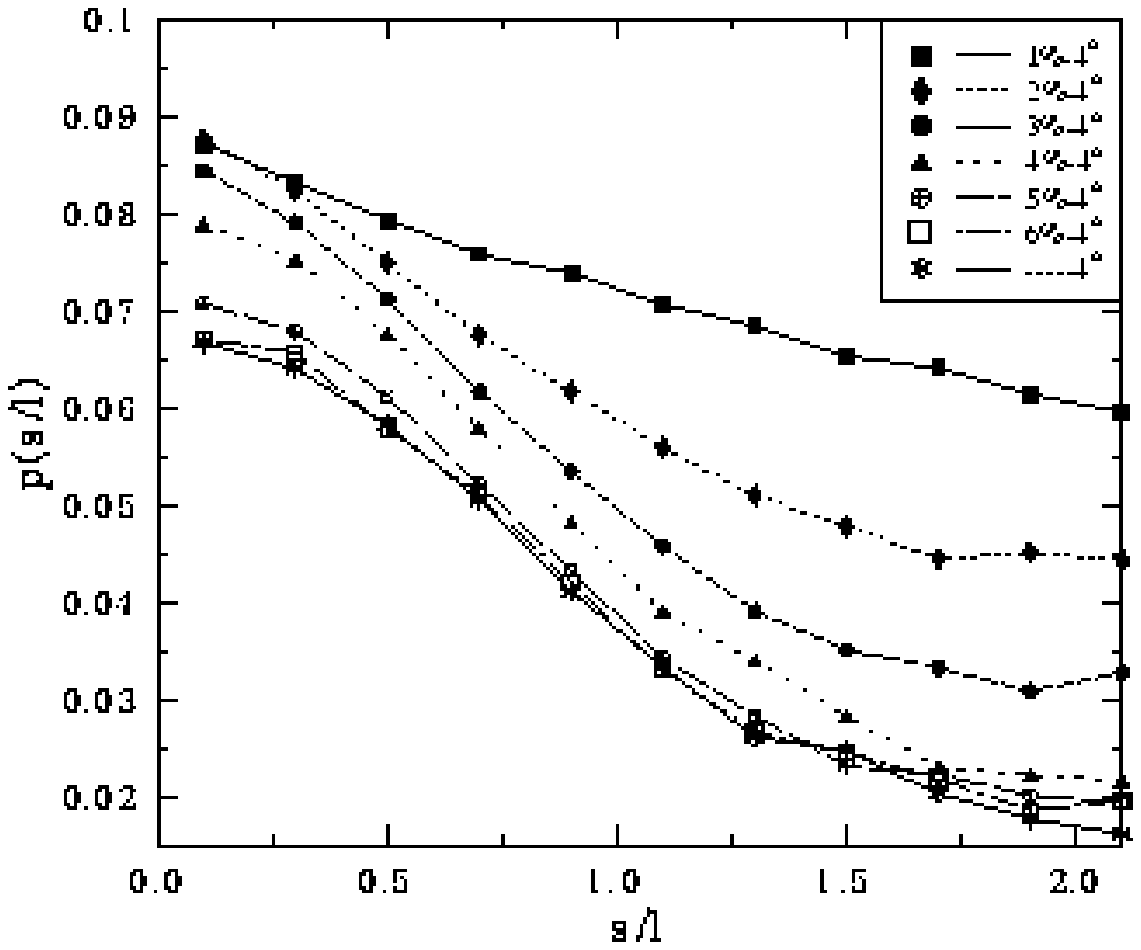,width=12cm}
\caption{The fraction of broken beams as a function of $s/l$
  varying the stretching threshold $t_{\epsilon} = 1\% - 6\%$. The fixed
  value of the bending threshold is $t_{\Theta} = 4^o$. For increasing
  $t_{\epsilon}$ the curves
  tend to a limit, which coincides with the result obtained by
  switching off the stretching mode.  }
\label{fig:prob_elo}
\end{center}
\end{figure}
\begin{figure}[!h]
\begin{center}
\epsfig{bbllx=117,bblly=374,bburx=501,bbury=686,file=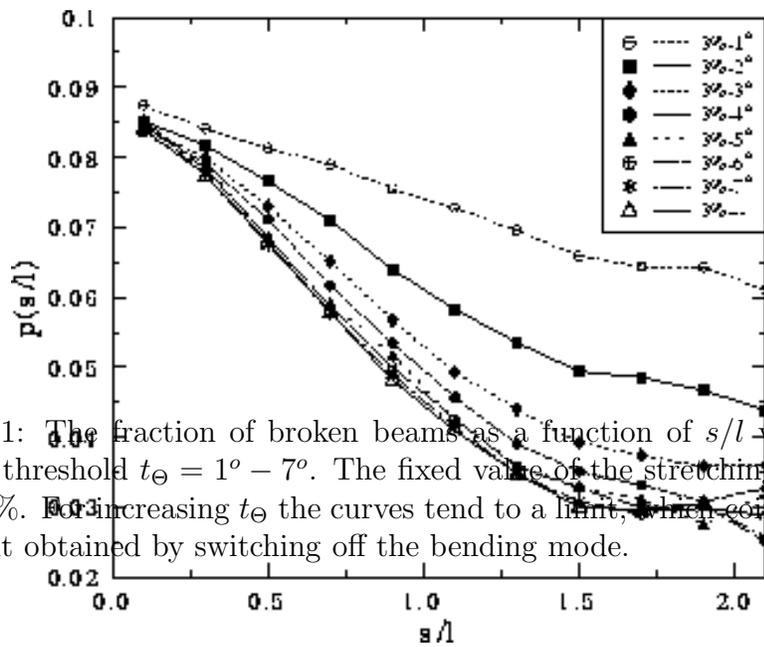,width=12cm}
\caption{The fraction of broken beams as a function of $s/l$
  varying the bending threshold $t_{\Theta} = 1^o - 7^o$. The fixed
  value of the stretching threshold is $t_{\epsilon} = 3\%$.
  For increasing $t_{\Theta}$ the curves tend to a limit,
  which coincides with the result obtained
  by switching off the bending mode.  }
\label{fig:prob_ang}
\end{center}
\end{figure}

\begin{figure}[!h]
\begin{center}
\epsfig{bbllx=117,bblly=374,bburx=501,bbury=686,file=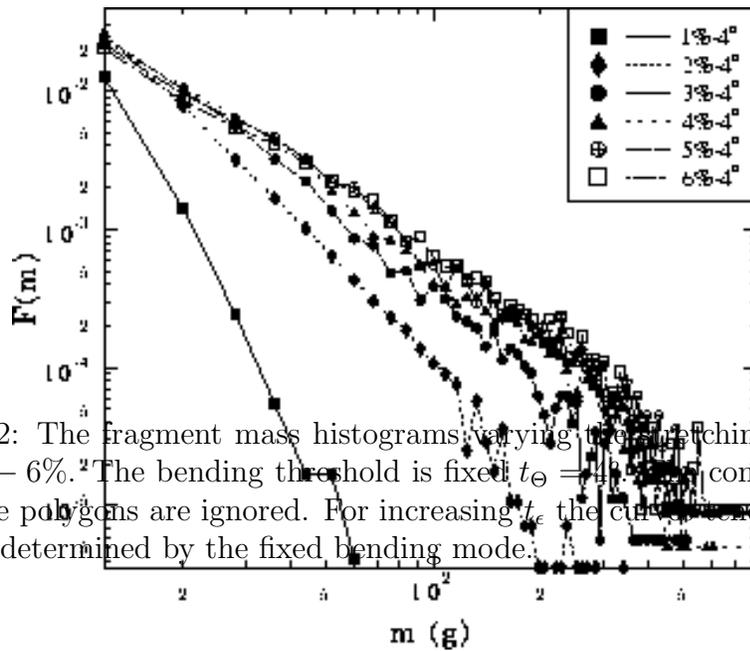,width=12cm}
\caption{The fragment mass histograms
  varying the stretching threshold $t_{\epsilon} = 1\% - 6\%$. The
  bending threshold is fixed $t_{\Theta} = 4^o$.
  The contribution
  of the single polygons are ignored. For increasing
  $t_{\epsilon}$ the curves tend to a limit, which is determined by the fixed
  bending mode.}
\label{fig:elo_dist}
\end{center}
\end{figure}
\begin{figure}[!h]
\begin{center}
\epsfig{bbllx=117,bblly=374,bburx=501,bbury=686,file=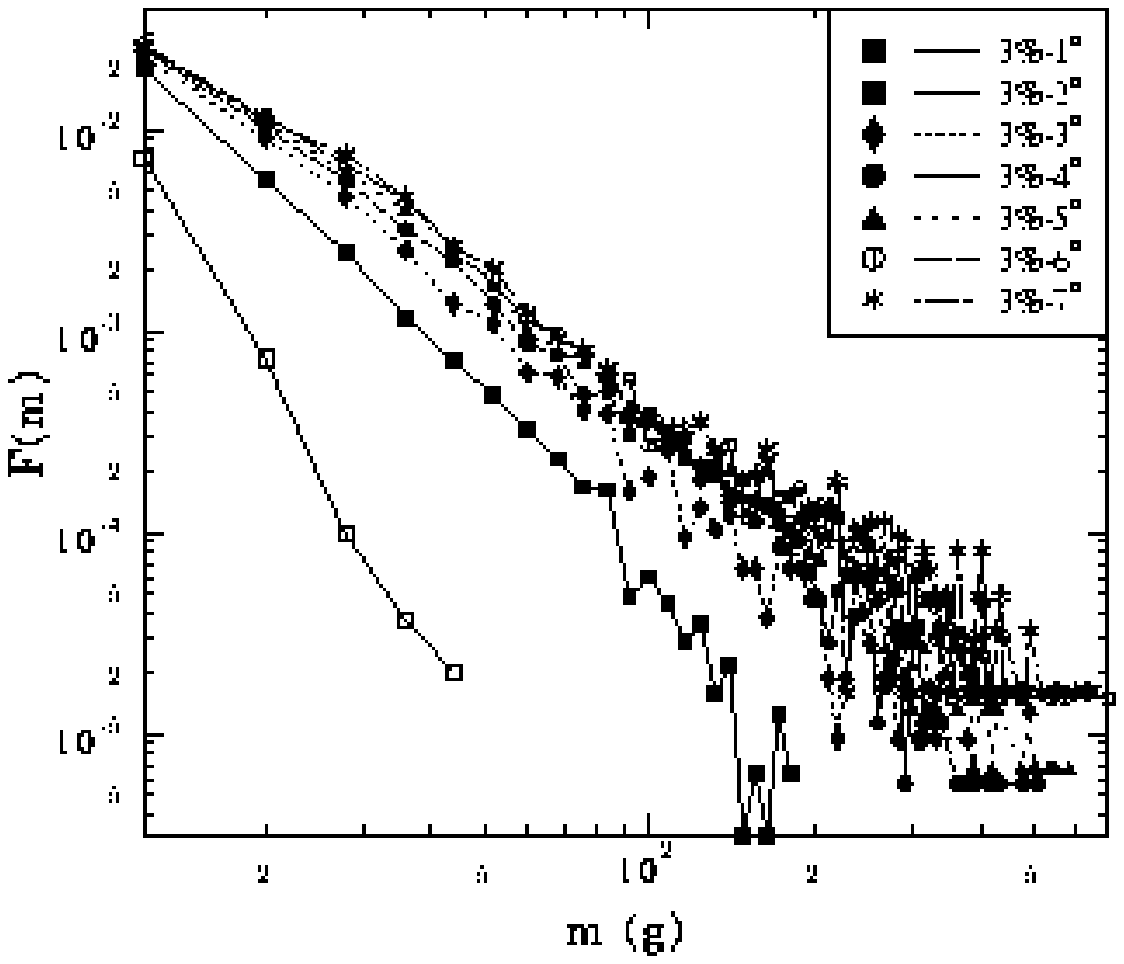,width=12cm}
\caption{The fragment mass histograms
  varying the bending threshold
  $t_{\Theta} = 1^o - 7^o$.
  The
  stretching threshold is fixed $t_{\epsilon} = 3\%$. The contribution
  of the single polygons are ignored. For increasing
  $t_{\Theta}$   the curves tend to a limit, which is determined by the fixed
  stretching mode.}
\label{fig:bend_dist}
\end{center}
\end{figure}

\begin{figure}[!h]
\begin{center}
\epsfig{bbllx=117,bblly=374,bburx=501,bbury=686,file=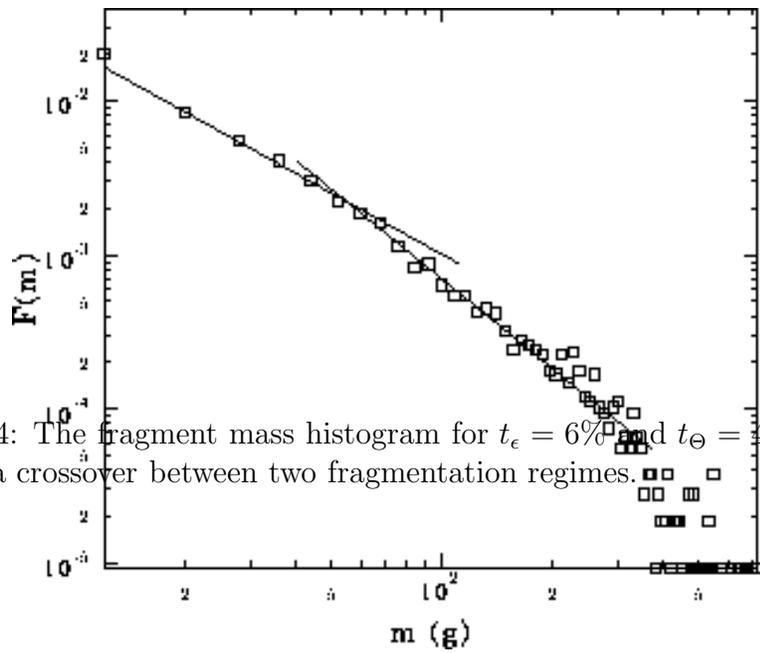,width=12cm}
\caption{The fragment mass histogram for $t_{\epsilon} = 6\%$ and
  $t_{\Theta} = 4^o.$ One can observe a crossover between
  two fragmentation  regimes. }
\label{fig:crossover}
\end{center}
\end{figure}
\begin{figure}[!h]
\begin{center}
\epsfig{bbllx=100,bblly=329,bburx=472,bbury=568,file=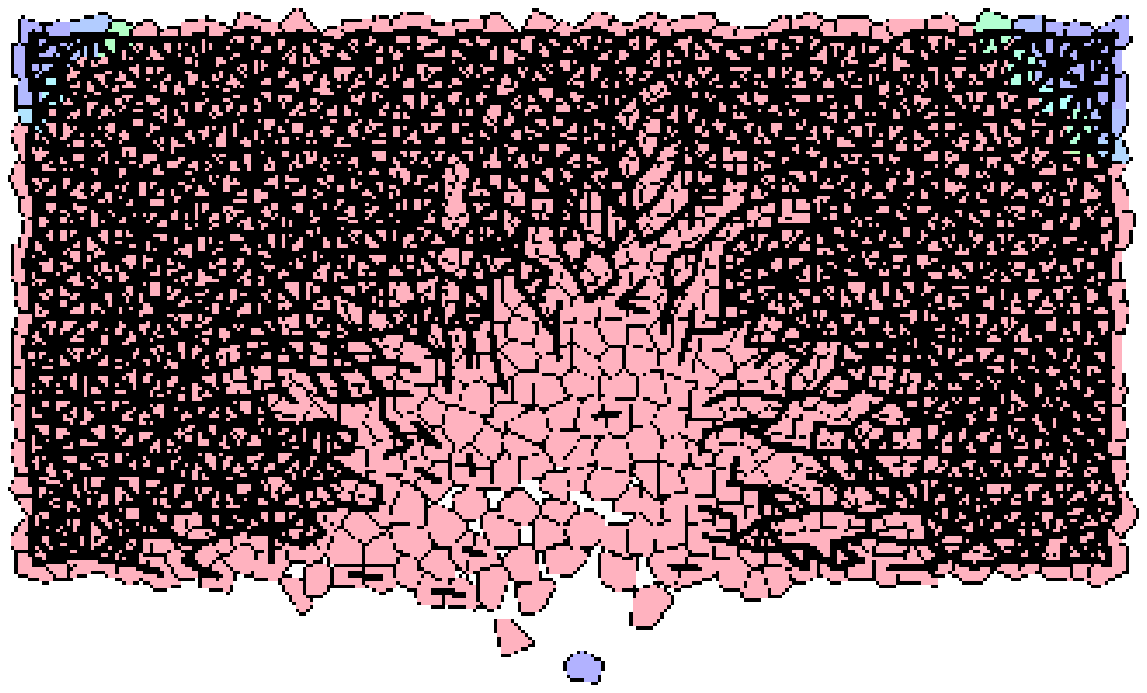,width=6.3cm}
\\
\epsfig{bbllx=88,bblly=278,bburx=502,bbury=618,file=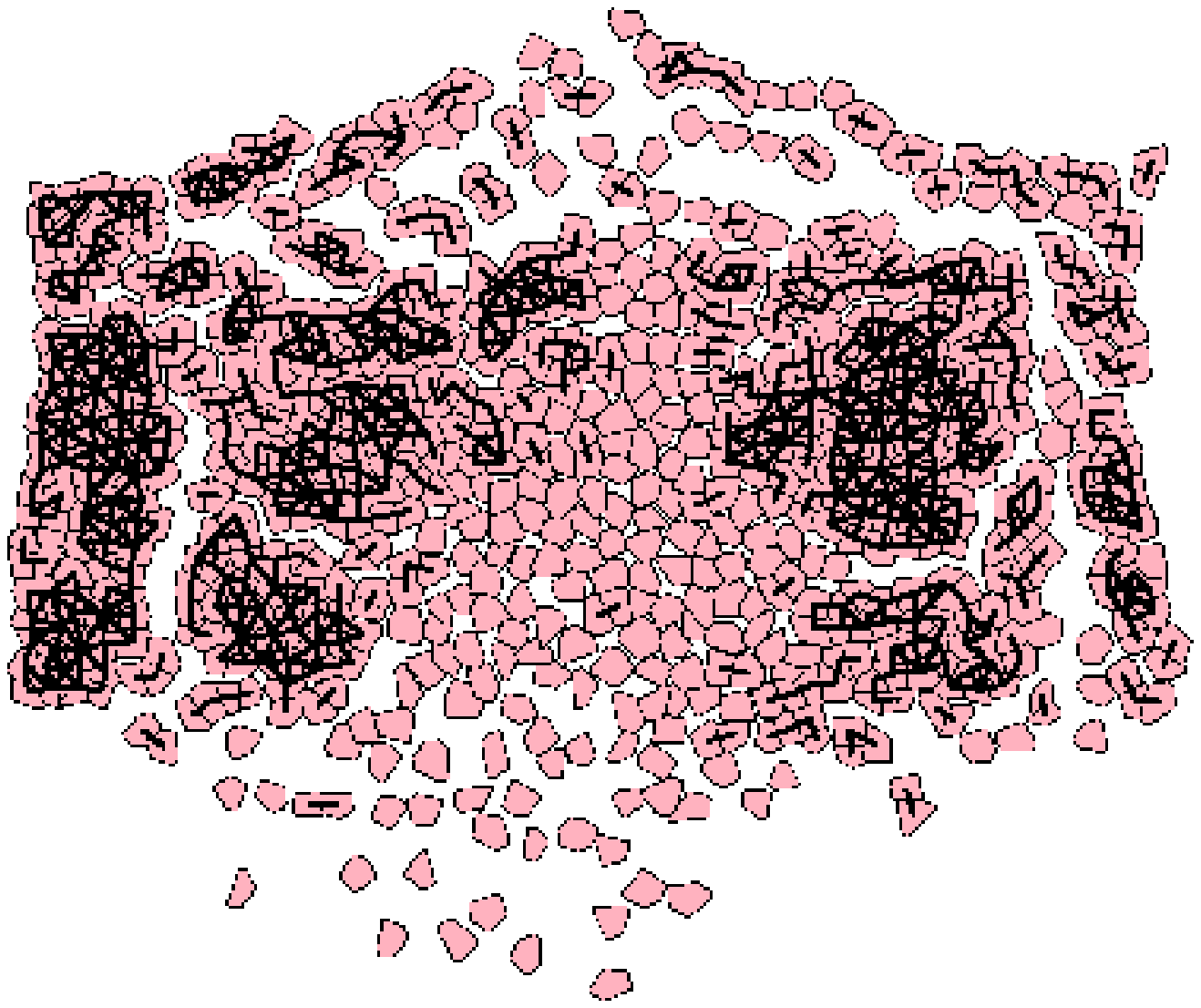,width=6.3cm}
\\
\epsfig{bbllx=66,bblly=231,bburx=538,bbury=700,file=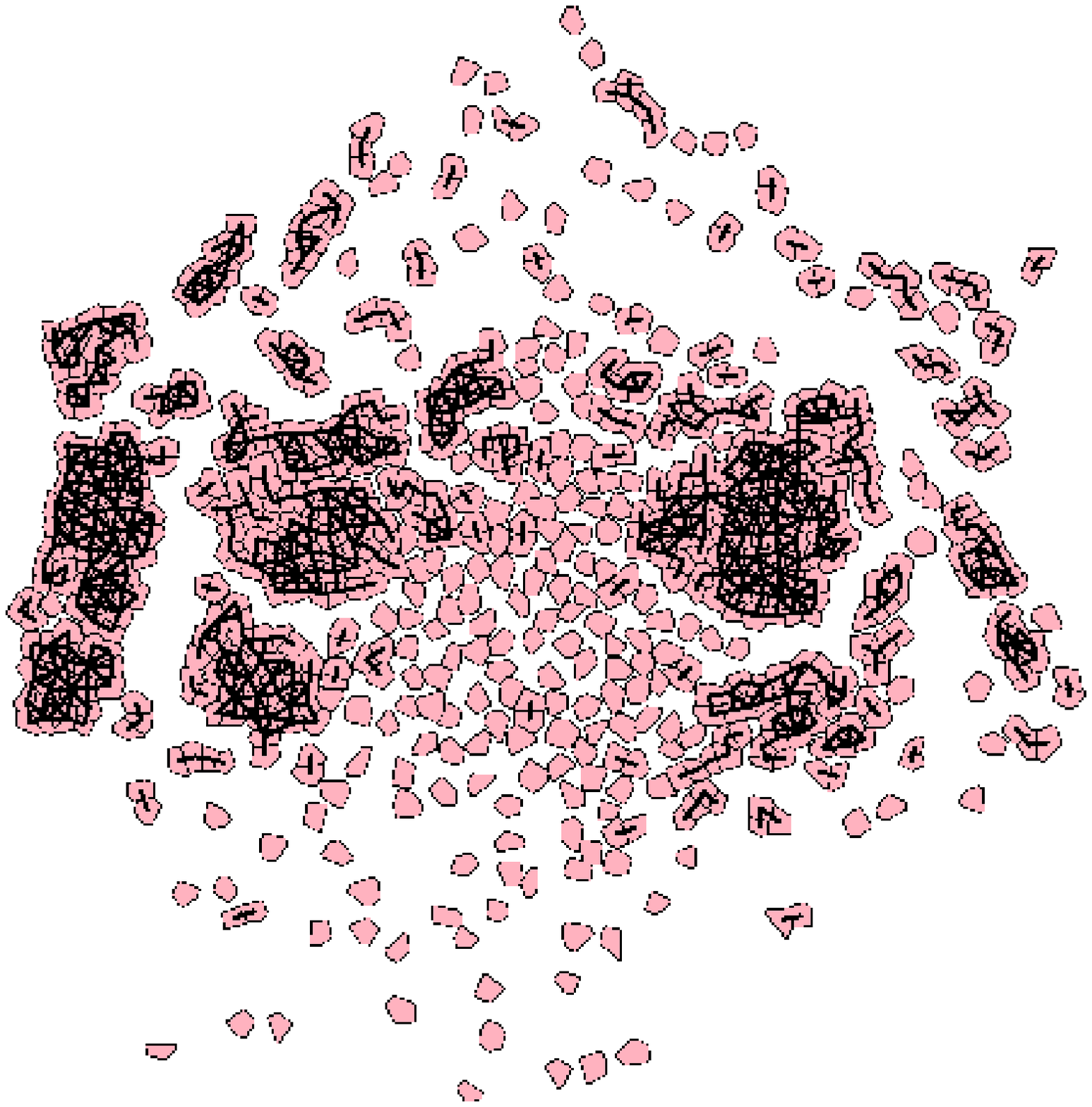,width=6.3cm}
\vspace{0.5cm}
\caption{Fragmentation of a concrete block composed of elastic grains.
One grain at the lower middle part of the block is given a high
velocity directed inside the block.  Here the velocity was $ 400 m/s$.
The size of the block was chosen to be $40 cm \times 20 cm$.
Snapshots of the evolving system are presented at
$ t= 0.0004 s$, $ t= 0.0015 s$ and
$ t= 0.003 s$.}
\label{fig:impact}
\end{center}

\end{figure}
\begin{figure}[!h]
\begin{center}
\epsfig{bbllx=117,bblly=374,bburx=501,bbury=686,file=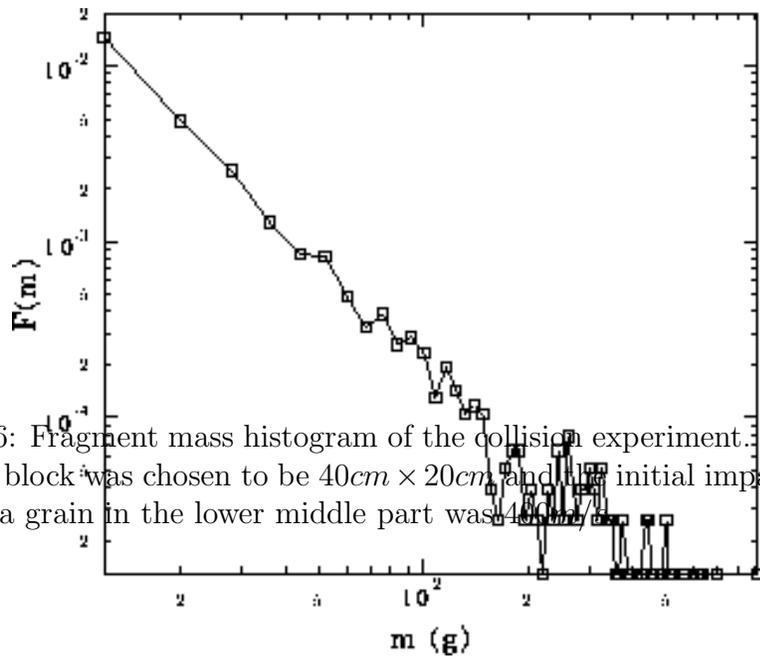,width=12cm}
\caption{Fragment mass histogram of the collision experiment. The size
  of the solid block was chosen to be $40 cm \times 20 cm$ and the
  initial impact
  velocity given to a grain in the lower middle part was $400 m/s$.}
\label{fig:coll_mass}
\end{center}
\end{figure}

\begin{table}[htbp]
\begin{center}
\vspace{0.4cm}
\begin{tabular}{|c|c|c|c|}
\hline
$Parameter$ & $Symbol$ & $Unit$ & $Value$  \\
\hline
\hline
Density & $\rho$ & $g/cm^{3}$ &5  \\
\hline
Grain bulk Young modulus & $Y$ & $dyn/cm^{2}$ & $10^{10}$ \\
\hline
Beam Young modulus & $E$  & $dyn/cm^{2}$ &$5 \cdot 10^9$  \\
\hline
Time step & $dt$  & $s$ & $10^{-6}$  \\
\hline
Diameter of the solid & $d$ & $cm$ & $40$ \\
\hline
Average \# of polygons & $n$ & & $1100$ \\
\hline
Energy of the explosion & $E_o$ & $erg$ & $5 \cdot 10^9$  \\
\hline
Average initial speed & $v_o$ & $m/s$ & $200$ \\
\hline
Estimated sound speed & $c$ & $m/s$ & $900$ \\
\hline
\end{tabular}
\caption{ The parameter values used in the simulations.}
\label{table_0}
\end{center}
\end{table}

\begin{table}[htbp]
\begin{minipage}[h]{\textwidth}
\begin{center}
\begin{tabular}{|c|c|c|c|}
\hline
\multicolumn{2}{|c|}{$t_{\Theta} = 4^o$}
&\multicolumn{2}{|c|}{$t_{\epsilon} = 3 \%$}  \\ \hline
$t_{\epsilon}$ & $\beta$ & $t_{\Theta}$ & $\beta$ \\ \hline
 1 $\%$ & (4.34) \footnote{These exponents
belong to the limiting case of extremely small
breaking thresholds. They were extracted
by fitting straight lines over the whole mass range of the
corresponding curves.}
 & $ 1^o $ & $(4.42)^\mathit{a}$  \\ \hline
 2 $\%$ & 2.39 & $ 2^o $ & 2.5   \\ \hline
 3 $\%$ & 2.01 & $ 3^o $ & 1.98  \\ \hline
 4 $\%$ & 1.97 & $ 4^o $ & 2.01  \\ \hline
 5 $\%$ & 1.95 & $ 5^o $ & 1.96  \\ \hline
 6 $\%$ & 1.96 & $ 6^o $ & 1.95  \\ \hline
        &      & $ 7^o $ & 1.97  \\ \hline
\end{tabular}
\caption{ The values of the exponent $\beta$ for different breaking
  thresholds.}
\label{table_1}
\end{center}
\end{minipage}
\end{table}

\end{document}